\documentclass[nonacm, sigconf]{acmart}
\setcopyright{none}
\renewcommand\footnotetextcopyrightpermission[1]{}
\usepackage{balance}
\usepackage{amsmath,amssymb,amsfonts}
\usepackage{graphicx}
\usepackage{textcomp}
\usepackage{xcolor}
\usepackage{array}
\usepackage{url}
\usepackage{multirow}
\usepackage{comment}
\usepackage[symbol]{footmisc}
\usepackage{mathtools}
\usepackage{siunitx}
\usepackage{booktabs}
\usepackage[linesnumbered,ruled,vlined]{algorithm2e}
\usepackage[noend]{algpseudocode}
\usepackage{mathtools}
\usepackage[symbol]{footmisc}\usepackage[final]{changes}
\definechangesauthor[name={Amirhossein Esmaili}, color=blue]{ae}
\definechangesauthor[name={Arash Fayyazi}, color=red]{af}
\usepackage{soul}
\usepackage{caption}
\usepackage{subcaption}
\SetKwRepeat{Do}{do}{while}%
\AtBeginDocument{%
  \providecommand\BibTeX{{%
    \normalfont B\kern-0.5em{\scshape i\kern-0.25em b}\kern-0.8em\TeX}}}

\setcopyright{acmcopyright}
\copyrightyear{2018}
\acmYear{2018}
\acmDOI{10.1145/1122445.1122456}

\acmConference[Woodstock '18]{Woodstock '18: ACM Symposium on Neural
  Gaze Detection}{June 03--05, 2018}{Woodstock, NY}
\acmBooktitle{Woodstock '18: ACM Symposium on Neural Gaze Detection,
  June 03--05, 2018, Woodstock, NY}
\acmPrice{15.00}
\acmISBN{978-1-4503-XXXX-X/18/06}

\begin{document}
\pagestyle{plain}

\title{HIPE-MAGIC: A Technology-Aware Synthesis and Mapping Flow for \underline{HI}ghly \underline{P}arallel \underline{E}xecution of \underline{M}emristor-\underline{A}ided Lo\underline{GIC}}


\author{Arash Fayyazi}
\authornote{Both authors contributed equally to this research.}
\affiliation{%
  \institution{University of Southern California}
  \city{Los Angeles}
  \state{California}
}
\email{fayyazi@usc.edu}

\author{Amirhossein Esmaili}
\authornotemark[1]
\affiliation{%
  \institution{University of Southern California}
  \city{Los Angeles}
  \state{California}
}
\email{esmailid@usc.edu}

\author{Massoud Pedram}
\affiliation{%
  \institution{University of Southern California}
  \city{Los Angeles}
  \state{California}
}
\email{pedram@usc.edu}


\begin{abstract}
\replaced[id=af]{Recent efforts for finding novel computing paradigms that meet today's design requirements have given rise to a new trend of processing-in-memory relying on non-volatile memories. In this paper, we present HIPE-MAGIC, a technology-aware synthesis and mapping flow for highly parallel execution of the memristor-based logic. Our framework is built upon two fundamental contributions: balancing techniques during the logic synthesis, mainly targeting benefits of the parallelism offered by memristive crossbar arrays (MCAs), and an efficient technology mapping framework to maximize the performance and area-efficiency of the memristor-based logic. Our experimental evaluations across several benchmark suites demonstrate the superior performance of HIPE-MAGIC in terms of throughput and energy efficiency compared to recently developed synthesis and mapping flows targeting MCAs, as well as the conventional CPU computing.}{
Exascale computing and its associated increasingly massive amounts of data storage and transfers from/to memory require increasingly efficient computing paradigms. Meanwhile, the PIM is increasingly gaining interest in both industry and academia as a serious contender to provide an efficient execution for memory-intensive applications. In particular, memristor, as a non-volatile memory element, has a theoretical built-in logic execution capability coupled with a significant reduction in latency and power when compared to the standard CMOS storage devices.}
\replaced[id=af]{}{The existing synthesis flows for arbitrary logic functions on a memristive crossbar array (MCA), do not usually consider the capability offered by target MCAs in terms of concurrent execution of operations. In this paper, we present some balancing techniques during the logic synthesis, mainly targeting benefits of the parallelism offered by MCAs and is further accompanied by an efficient technology mapping framework to maximize the performance and area-efficiency. Our experimental evaluations across ISCAS'85 and IWLS'93 combinational circuit benchmark suites demonstrate that the HIPE-MAGIC can significantly improve both the average computational latency (2.12x) and area (1.38x).}
\end{abstract}

\maketitle

\section{Introduction}
In a classical \textit{von Neumann} architecture, the processor and memory are separate entities, and thus there is the need for the data transfers between them to perform a computational task. \replaced[id=ae]{These data transfers lead to performance degradation and energy consumption\cite{seshadri2017ambit,seshadri2014willow,fujiki2018memory}.}{\cite{seshadri2014willow,seshadri2017ambit,fujiki2018memory}As advances in processor speeds have greatly exceeded those in memory speeds over the last few decades, this separation has resulted in a major performance bottleneck in modern high performance computing systems.} The situation is exacerbated because of the limited cache capacity and on-chip bandwidth\cite{balasubramonian2014near}, which give rise to long idle time intervals due to the memory synchronization during a computation\cite{wulf1995hitting}. The processing-in-memory (PIM) approach attempts to address this issue by breaking the von Neumann separation, and eliminating the need to transfer data between the processor and memory, since data already resides in the memory \cite{floatPIM}.
Emergence of new non-volatile memory technologies such as memristors which can be employed to perform logic operations, has provided new avenues for employing PIM. Memristors offer relatively high endurance, switching speeds of less than 10 ns, and data retention of about 7 years \cite{MemristorChar}. One approach for performing logic computations within a conventional memristive crossbar array (MCA) is to use a \textit{stateful logic} design style. In the stateful logic, logic states are represented in terms of variable resistance values of memristors. Typically, the high resistance state of a memristor is used for presenting logic 0 whereas the low resistance state is used for presenting logic 1. IMPLY \cite{borghetti2010memristive} and MAGIC \cite{MagicBenefits} are two examples of widely used stateful logic design styles. 

There has been a body of research on developing an automatic framework for the efficient implementation of arbitrary logic functions within a MCA \textcolor{black}{based on the IMPLY \cite{kvatinsky2011memristor, Imply, BDDSyn} or MAGIC design style \cite{SimAnn,Magic, Said,Simpler}}. \deleted[id=ae]{Existing works based on IMPLY logic perform logic computations in the crossbar array by applying a sequence of material implication operations. These operations should usually be performed on a single row of the crossbar.}The MAGIC-based implementation requires memristors only whereas the IMPLY-based implementation needs both memristors and resistors to realize the material implication function. Furthermore, IMPLY-based solutions rely on some custom data structures which are not typically part of a standard synthesis flow, while MAGIC-based implementations use the conventional logic NOR gate. In addition, the NOR gates in the MAGIC design style \deleted[id=ae]{, however,}can be implemented in the crossbar alongside both rows and columns.
This allows parallel operations of several NOR gates, which in turn tends to decrease the computation latency. 

\deleted[id=ae]{Furthermore,} The traditional flow for implementing logic functions within a MCA consists of two phases: logic optimization and technology mapping. Previous work typically employ a standard synthesis tool, such as the ABC tool \cite{ABC}, to obtain a gate-level netlist, and focus more on the second phase which is assigning suitable memristors on the MCA to logic gates in order to  minimize  delay and/or area while satisfying the required \textit{spatial alignment} constraints for memristors within the MCA.  This approach does not result in an efficient logic implementation on the MCA because it does not consider the advantages that the underlying 2D grid of the MCA can offer in terms of parallel computation of logic operations. 

This paper presents HIPE-MAGIC, a synthesis and mapping flow which exploits the native characteristics of MCA. All in-memory computations in this work rely on the MAGIC style of computing.
The main contributions of this work are four-fold: 
\begin{itemize}
    \item A look-up table (LUT)-based synthesis flow (where each LUT groups several MAGIC NOR gates), in which the flow is accompanied by balancing operations that provide more opportunities for parallelization of computations in the underlying MCA by reducing the logic depth of the target circuit to be mapped to MCA.
    \item A heuristic mapping strategy which helps improve both the the area efficiency and computation latency of the MCA, realizing an LUT-based logic network.
    \item Improving state-of-the-art synthesis and mapping flows targeting MCAs. More precisely, HIPE-MAGIC improves both the average computation latency (by a factor of 2.1) and area (by a factor of 1.4) across ISCAS'85 and IWLS'93 combinational circuit benchmark suites compared to prior work.
    \item A detailed analysis of strengths and weaknesses of our PIM solution compared to the conventional CPU computing (as an example of von Neumann architecture), based on a parameterized analytical model. The analysis demonstrates that PIM-based solution generated by HIPE-MAGIC have superior performance in terms of throughput and energy efficiency compared to those generated by recently developed synthesis and mapping flows, as well as the conventional CPU. 
\end{itemize}

While the scope of this work is focused on  a synthesis and mapping flow in PIM paradigm and not comparing the PIM approach with the traditional CPU computing, we also present a analytical comparison between the two in Section \ref{analytical_compare} in terms of throughput and energy efficiency based on a parameterized analytical model. The remainder of the paper is organized as follows. \deleted[id=ae]{Sec. \ref{PriorWork} provides a survey on prior work whereas}Sec. \ref{PreMot}  describes central preliminary concepts used in this work. Details on the newly proposed synthesis and mapping frameworks of HIPE-MAGIC are presented in Sec. \ref{Flow}, while Sec. \ref{results} studies the effectiveness of HIPE-MAGIC. Conclusions are drawn in Sec. \ref{conc}.

\section{Background} \label{PreMot}

All in-memory computations in this work rely on the basic MAGIC NOR operation. \added[id=af]{We first describe two phases of logic synthesis, technology-independent and technology-dependent phases. Then, we} explain MAGIC style operation and supergate-aided synthesis and mapping techniques,
where the latter drives our choice for the proposed flow in this work. Finally, we explain how MCA motivates our choice to optimize both phases of the logic synthesis.
\subsection{Preliminaries}

\subsubsection{Logic Synthesis}
Logic synthesis is divided into technology-independent and technology-dependent phases. In the first phase, algebraic transformations are performed to reduce the number of literals in the \textit{optimal} factored form of a given Boolean expression and consequently reduce its area and delay. These algebraic optimizations are interleaved with node simplification operations to utilize controllability and observability don't cares and thereby achieve even more reduction in area or delay of the synthesized network.  Next, in preparation for the technology-dependent phase, the optimized Boolean network is transformed into a common semantic domain e.g., an And Inverter Graph (AIG). Note that the library cells are also represented as pattern graphs in the same domain. This step is called technology decomposition, and such a network is called a subject graph. Finally, as the technology-dependent phase, subgraphs of the subject graph are mapped to suitable pattern graphs (e.g., gates from library cells, or LUTs for FPGAs) in order to minimize the target cost function, e.g, delay or area. This is a graph covering problem.

\subsubsection{MAGIC Style}
\begin{figure}[!t]
\centering
\includegraphics[scale = 1.0]{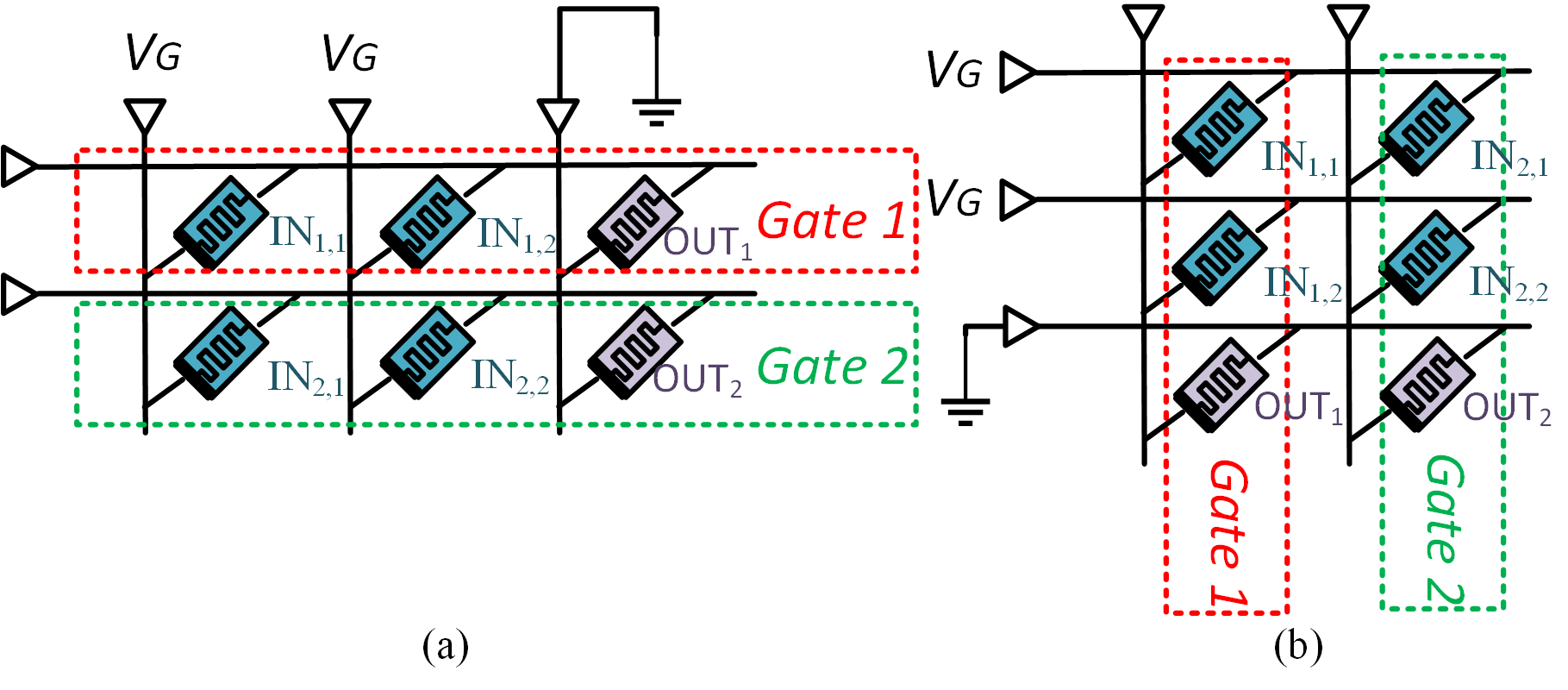}
\caption{\small Parallel execution of two aligned MAGIC NOR gates in a) row-wise and (b) column-wise manners.}
\label{fig:Magic}
\end{figure}

The MAGIC NOR function is performed by applying a single operating voltage ($V_G$) to the input memristor(s) to initialize  the output memristor to logic 0 (high resistance value).  The state of the output memristor (which is also considered as a memory cell) changes pursuant to the logical states of the input memristor(s). The NOT function is realized as a single-input NOR function.  The spatial requirement for a MAGIC gate is that its input(s) and output must be located on the same row, or in the same column of the memristor array. Furthermore, to execute multiple row-wise (or column-wise) MAGIC gates in parallel, their input and output memristors should be aligned along the same set of columns (rows). Fig. \ref{fig:Magic} depicts crossbar configurations for in-memory execution of two row-wise and two column-wise MAGIC NOR gates in parallel. This figure shows the advantages of MAGIC operation such as the separation between the input(s) and output memristors, the need for only a single $V_G$, absence of additional peripheral elements, and the ability to perform multiple operations in parallel \cite{Magic}. 
\added[id=ae]{Since NOR is a complete logic function, a MAGIC NOR gate is sufficient for the execution of any logic operation within the MCA.}

\newcommand\widebar[1]{\mathop{\overline{#1}}}

\subsubsection{Supergate-Aided Synthesis and SoP-to-NoN Translation} \label{sec.said}
Tenace et al. \cite{Said} proposed to use $k$-input LUTs \added[id=ae]{for mapping the subgraphs of an AIG subject graph to the MCA}. In their approach, these LUTs are subsequently transformed to NOR-of-NORs (NoN) \textit{supergates} and implemented on the MCA. Any single-output Boolean function $\mathcal{F}: \mathbb{B}^n \to \mathbb{B}$ represented by a Sum of Products (SOP) \added[id=ae]{associated with a single LUT} can be translated to NoN following these three steps: 1) negating all primary inputs, 2) replacing all AND and OR operations ($\wedge$ and $\vee$) with NOR operations. ($\widebar{\mathbb{\vee}}$), and 3) negating the result.
The crossbar mapping scheme proposed in \cite{Said} for these supergates is a tile-based mapping, where each tile contains a supergate and is expressed by two indices $l$ and $y$ representing the column and row coordinates of that tile, adhering to some spatial rules. 
All supergates from one logic level are placed in one column and sorted by their size (i.e., the number of their LUT terms). The mapping scheme in \cite{Said} also imposes a constraint that a supergate in $(l,y)$ position must be smaller in size compared to the supergate placed in $(l-1,y)$ tile. 

\begin{figure}[!t]
\centering
\includegraphics[scale = 0.82]{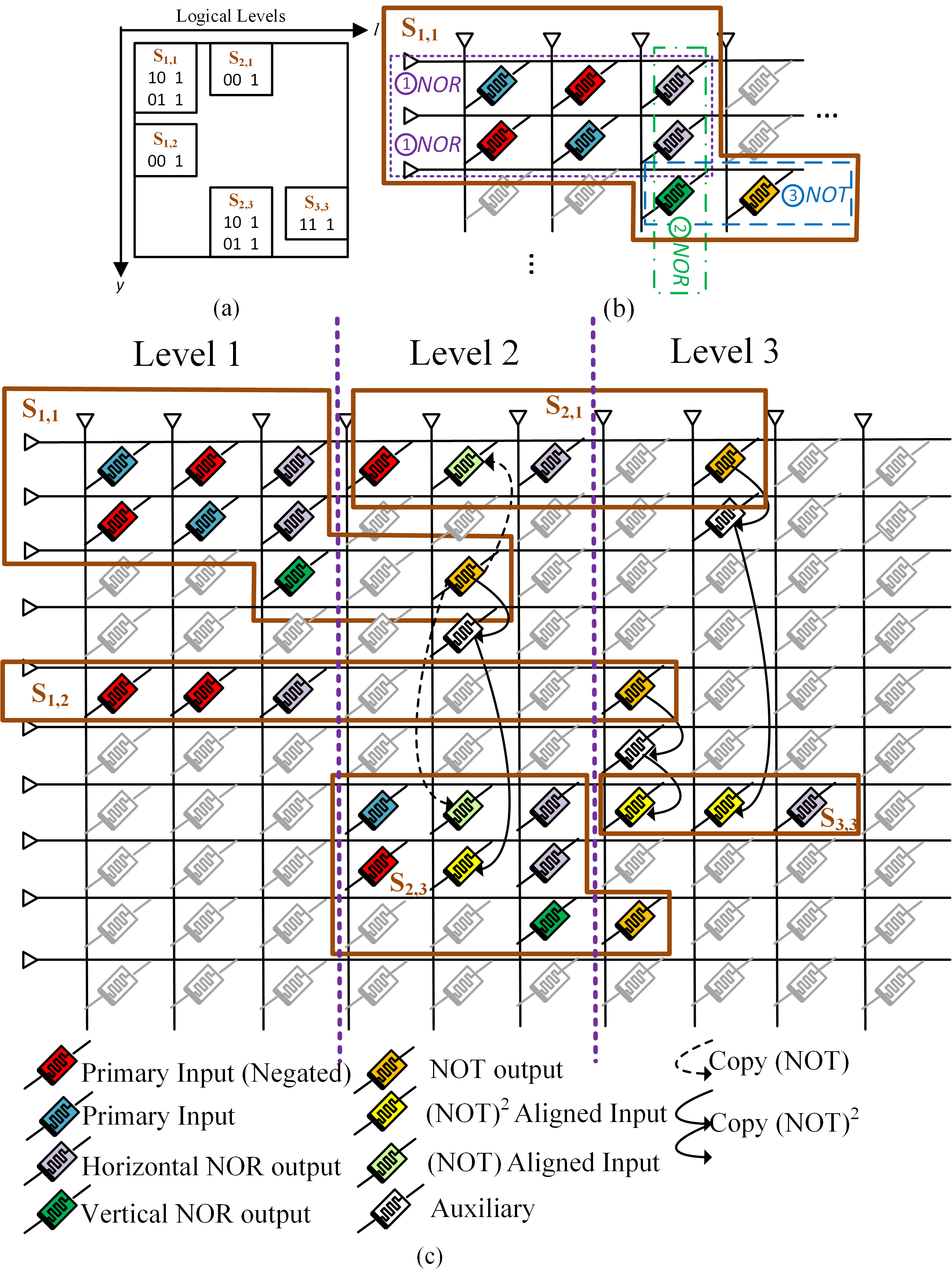}
\caption{\small 
Implemented 1-bit full adder based on \cite{Said}: (a) tile-based mapping, (b) sequence of operations within a supergate, and (c) MCA mapping. 17 cycles of PIM operations and 30 memristors are needed. \textcolor{black}{ Copy(NOT) takes one cycle, whereas Copy(NOT)$^2$ takes two cycles of PIM} operations.
}
\label{fig:SAID}
\end{figure}

An example of the tile-based mapping for a single-bit full adder is depicted in Fig. \ref{fig:SAID} (a). An output of a supergate can be generated as a sequence of three MAGIC functions as suggested in Fig. \ref{fig:SAID} (b). To cascade supergates, alignment copy operations are performed to satisfy the spatial constraints of MAGIC gates. Depending on the polarity of input/output signals, the data alignment copy operations are either NOT copies, or (NOT)$^2$ copies (see Fig. \ref{fig:SAID} (c)). (NOT)$^2$ copy represent two consecutive NOT copies which use one extra auxiliary memristor for the intermediate signal. 

\textcolor{black}{A major challenge of prior work including and \cite{Said} and \cite{SimAnn} is the large computational latency due to alignment copy operations. For instance, main logic operations in Fig. \ref{fig:SAID} takes only five cycles while alignment copy operations takes 12 cycles. HIPE-MAGIC tackles this challenge by devising a mapping scheme that can further reduce operational computation by overlapping input/output signals as well as sharing alignment copy operations leveraging heuristic placements.}


\subsection{Motivational Example for MCA-Driven Optimization}

Previous work on optimizing synthesis flow for CMOS illustrate the effect of the \textit{balancing operation} that reduces the logic depth of a combinational circuit on reducing its delay. In \cite{SOPBalance}, Michenko et al. proposed AND balancing and SOP balancing algorithms for the delay optimization. AND-balancing of an AIG is a well-known fast transformation that reduces the number of AIG levels. Pasandi and Pedram \cite{Balancing} also presented balanced factorization and rewriting algorithms by considering an imbalance factor when calculating the value of a potential factorization.  Their results show the importance of the AIG balancing operation in minimizing the delay.

\begin{figure}[t]
\centering
\includegraphics[scale = 0.9]{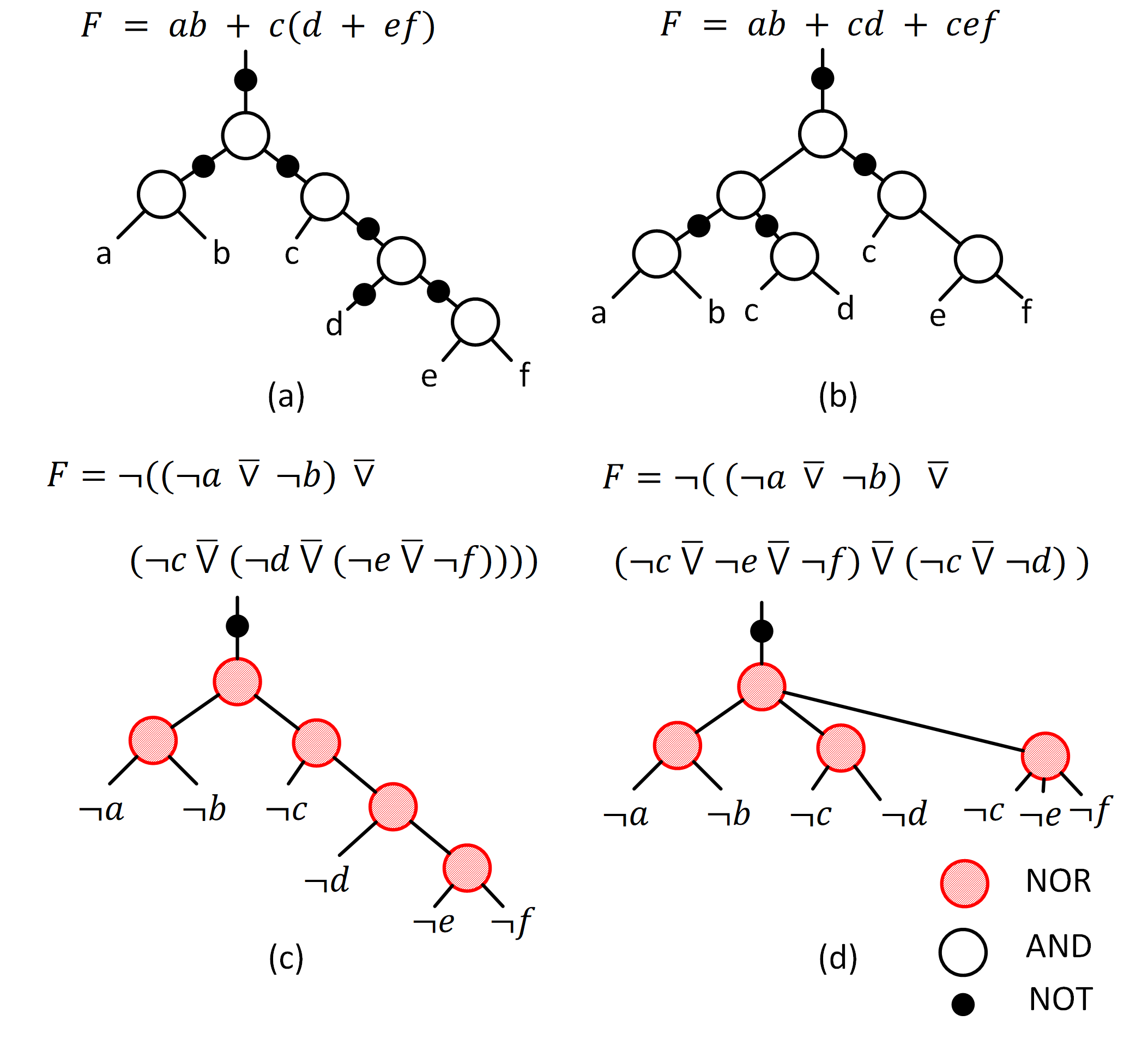}
\caption{\small An example of SOP balancing: (a) an unoptimized and (b) optimized implementation in AIG, which can be translated to (c) an unoptimized and (d) optimized implementation in MAGIC style, respectively.}
\label{fig:Balance}
\end{figure}

To perform the SOP balancing, a small AIG (e.g., an AIG that depends on say 10 or fewer inputs)  is converted into a SOP. Next, the SOP balancing applies AND-balancing to each product and subsequently to the sum operation.
Fig. \ref{fig:Balance} illustrates the SOP-balancing technique on a small AIG (inspired by \cite{SOPBalance}), and how this can be translated into delay optimization of the logic implementation of MAGIC NORs. The total computational cycles needed for implementing this function on the memristive crossbar (e.g., number of logic levels) is reduced from 5 in Fig. \ref{fig:Balance} (c) to 3 in Fig. \ref{fig:Balance} (d). This example motivates us to leverage balancing operations, which can lead to improvements for MCA-based PIM by enabling more operations to be performed in parallel.

\section{HIPE-MAGIC's flow} \label{Flow}
\textcolor{black}{HIPE-MAGIC's flow} is divided into two main steps:
\begin{itemize}
    \item \textbf{Logic Synthesis:} The input is an arbitrary logic function, and the output is a netlist comprising optimized mapped LUT representation of the input function.
    \item \textbf{MCA Mapping:} The input is a netlist of synthesized LUTs, and the output is the placement of each supergate within the 2D memristive crossbar array along with a schedule of the sequence of operations.
\end{itemize}

\subsection{Technology-aware Logic Synthesis}\label{AA}
Logic synthesis for enabling PIM is a crucial step in the MCA design flow with a significant impact on the total area and performance. 
The underlying 2D grid in MCAs offers the opportunity for concurrently performing many logic operations in each logic level. This opportunity should be considered during the logic synthesis flow.
Similar to prior work, we utilize the ABC tool \cite{ABC}, but with the addition of balancing operations. ABC receives an arbitrary logic function, and generates a netlist of optimized LUTs while minimizing the number of gates and logic levels. HIPE-MAGIC runs balancing operations in two phases within the ABC tool: 
\begin{enumerate}
    \item \textit{Technology-independent phase:} on the AIG itself (i.e., before mapping the Boolean function to LUTs).
    \item \textit{Technology-dependent phase:} on the priority cuts (i.e., during the LUT mapping process).
\end{enumerate} 

\textbf{Technology-independent Phase:}
We have added balanced factorization and rewriting algorithms \cite{Balancing} as commands to ABC \cite{ABC}. These commands tend to reduce the size of the AIG representing the input logic network, and also introduce more parallelism in the AIG by utilizing the balancing operations. We have used 
the following optimization scripts: \textit{\{blnc\_syn2, resyn, resyn2, resyn2rs, compress2rs\}}  
where blnc\_syn2 is a set of optimization commands, which is interleaved with balanced rewriting and factorization commands together with the AND balancing command of ABC. The other commands are heuristics methods, provided by the ABC synthesis to further optimize the AIG network. 

\textbf{Technology-dependent Phase:} In general, AND-balancing, balanced factorization, and rewriting algorithms are limited to AIG and AND gates, while SOP-balancing can look at larger functions (i.e., the priority-cuts which are used for the $k$-input LUT mapper in ABC \cite{ABC}, where $k$ denotes the number of LUT inputs). Therefore, SOP-balancing can be applied during the LUT mapping to  reduce delay in many cases. However, it is impossible to apply SOP-balancing algorithms to each of the AIG nodes since large design can have millions of AIG nodes. For this purpose, a large AIG is divided  into parts (using cutlines), and the SOP-balancing operation is carried out for each of the cuts (for an example, see Fig. \ref{fig:Balance}).

The SOP-balancing operation is implemented in ABC, and called through the priority-cut based mapper \textit{if} command. We called our optimization script \textit{TechDepOpt}, which performs the technology-dependent synthesis targeting MCA. The sequence of ABC commands are: 
\textit{st; if -K \textbf{k}; (st; dch; if -K \textbf{k}$)^2$\footnote[2]{The number of rounds of running the sequence of the commands in parenthesis.} ; st; if -g -C \textbf{c} -K \textbf{l}; (st; dch; if -K \textbf{k}$)^2$; st; dch; if -K \textbf{k} -S \textbf{s}.} In the \textit{TechDepOpt}, \textit{-g} enables SOP-balancing for the cut evaluation and \textit{c} is the number of \textit{l}-input cuts computed at each node in the subject graph. \textit{dch} performs the AIG-based synthesis with a repeated sequence of technology-independent optimizations on different structural choices (which are functionally equivalent networks obtained by running AIG rewriting scripts on the current network), and \textit{st} transforms the network back to the AIG form. To ensure that the delay after mapping into \textit{k}-LUT can be reduced, the cut size for SOP-balancing operation (\textit{\textbf{l}} in "\textit{if -g -C \textbf{c} -K \textbf{l}}") is set to a number larger than the cut size for the final mapping command (\textit{\textbf{k}} in "\textit{if -K \textbf{k} -S \textbf{s}}").
Using a larger cut size for SOP balancing operation can potentially result in better delay optimization. 

\subsection{Technology Mapping} \label{sec.mapping}
The method presented in \cite{Said}, for mapping a netlist of supergates to the MCA, tends to limit the potential performance gain that can be obtained by the parallelization potential offered by the supergates. It also introduces a number of redundant operations, according to how it employs the SOP-to-NoN translation, which can be avoided. \textcolor{black}{Furthermore, to make better use of the 2D crossbar grid, the area-efficiency of mapped gates, associated with how densly memristors used for logic computations are placed next to each other in rows and columns, is preferable.} 
However, the mapping rules presented in \cite{Said} may result in  a sparse mapping of a netlist (e.g., see Fig. \ref{fig:SAID}), although in fact  a denser and more area-efficient mapping may be possible. In addition, the overhead associated with the data alignment copies can be alleviated by reusing auxiliary memristors.
Based on above observations, certain refinements and mapping operations are used  as shown in Algorithm \ref{alg.mapping} and elaborated on below.

\begin{figure}[t]
\centering
\includegraphics[scale = 0.9]{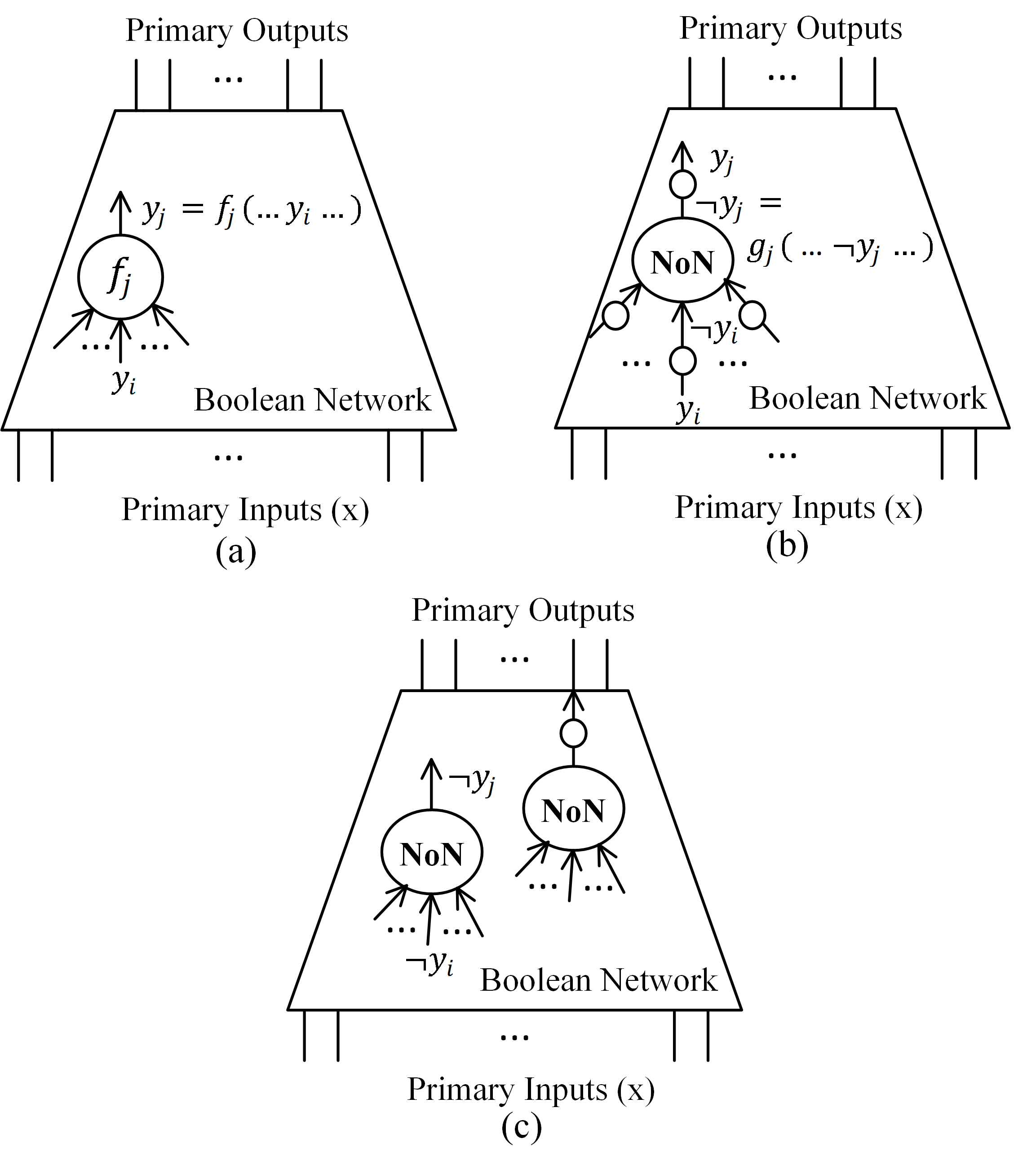}
\caption{\small An example of MCA-driven SOP-to-NoN translation: (a) $f_j$ is a logic function (represented in the SOP form), (b) original translation to NoN where $g_j$ is a logic function composed of only NOR functions, and (c) refined translation. Small circles denote the NOT function.}
\label{fig:ref1}
\end{figure}

\textbf{MCA-Driven SOP-to-NoN Translation:} For each pair of cascaded LUTs, we merge the third step of the SoP-to-NoN mapping of the second LUT, cf. Section \ref{sec.said} above, with the first step of the SoP-to-NoN mapping of the first LUT (this is done as part of \textit{Map\_SOP\_to\_NoN()} function in Algorithm \ref{alg.mapping}). 
For instance, assume a Boolean network which is a graph of connected blocks (e.g.,  nodes, LUTs). Here, the individual component blocks are 2-level Boolean functions in the SOP form, which can be represented by $y_j = f_j (x,y)$ where $y_j$ denotes the node's output variable and $x$ and $y$ denote node's input variables, respectively (see Fig. \ref{fig:ref1} (a)). There exists an edge $(i,j)$ if $f_j$ depends explicitly on $y_i$; differently worded, there is a connection between the $j^{th}$ node and the $i^{th}$ node. The SOP-to-NoN translation imposes that each supergate (e.g., the node that is created after translation) depends on negated value of its input nodes, $g_j(\neg x,\neg y)$ as shown in Fig. \ref{fig:ref1} (b) where $y_j = \neg g_j(...,\neg y_i,...)$. Therefore, $y_i$ is calculated by performing a NOT function on $\neg y_i$ which is the node variable of $g_i$. This NOT function is equivalent to the last step in the translation of the SOP representation of $f_i$ into the NoN representation. Eliminating the two NOT functions on an edge between two cascaded nodes (see Fig. \ref{fig:ref1} (c)) reduces the number of operations required for implementing the target Boolean network. In other words, for each supergate (except for the ones that produce the primary outputs), we discard the third operation (e.g., the NOT function in Fig. \ref{fig:SAID} (b)). 

\let\oldnl\nl
\newcommand{\nonl}{\renewcommand{\nl}{\let\nl\oldnl}}
\begin{algorithm}[h]
\SetAlgoLined
\KwIn{The synthesized netlist of LUTs\\ \hspace{30pt}$L$: Number of logic levels}
\KwOut{Mapping of supergates and alignment copies}
\nonl // MCA-Driven  SOP-to-NoN  Mapping:\\
Map\_SOP\_to\_NoN()\;

\nonl //  Placement of supergates:\\
\For{$l = 1;\ l \le L;\ l = l + 1$}{Sort\_Supergates($l$); // sort supergates in level $l$ \\\If{($l\mod2==1)$}{Flip\_Supergates($l$); // flip supergates in level $l$ vertically}Place\_Supergates;($l$) // place supergates in level $l$}
\nonl // Resource Sharing for Data Alignment Copies:\\
Share\_Align\_Memristors()\;
    \caption{The proposed mapping scheme}
    \label{alg.mapping}
\end{algorithm}

\textbf{Placement of Supergates:}
The spatial rules for the mapping procedure mentioned in Section \ref{sec.said}, rely on the intuition that as the logic level increases, the supergate size decreases. However, this intuition does not always hold true. For instance, in the mapping of the one-bit full adder depicted in Fig. \ref{fig:SAID}, $S_{3,3}$ and $S_{2,3}$ will be pushed down in each column (logic level) following the spatial rules although they can be placed higher up in those columns.
These replacements increase the area-efficiency of the mapped supergates, which thus make better use of the underlying 2D grid of memristors.

We first sort the supergates in each logic level ($l$) in the decreasing order of their size ((\textit{Sort\_Supergates($l$)} function in Algorithm \ref{alg.mapping})). To take advantage of the underlying 2D grid, we alter the direction of the vertical NOR in supergates for consecutive columns (this is done in \textit{Flip\_Supergates($l$)} function in Algorithm \ref{alg.mapping}). Next, we place the supergates of first logic level ($l=1$) in the first column following the sorting order. For subsequent columns, we go through the sorted supergates in each level and place each supergate in the first available vertical position ($y$) in the column where the output of the supergate in the same $y$ and level $l-1$ would be one of the inputs of the supergate in level $l$. We also align the supergate in level $l$ with the one in level $l-1$ in a way that the memristor corresponding to the output of the supergate in level $l-1$ is shared with the corresponding input of the supergate in level $l$ in its first row. In this way, one data alignment copy is saved for each pair of cascading supergates in the same $y$. At the end, for those supergates in level $l$ that were not placed in column $l$ as a result of the described process, we place them in the decreasing order of their size in the first available $y$ in column $l$ (this is done in  \textit{Place\_Supergates($l$)} function in Algorithm \ref{alg.mapping} for each column $l$). 

\begin{figure}[t]
\centering
\includegraphics[scale = 0.85]{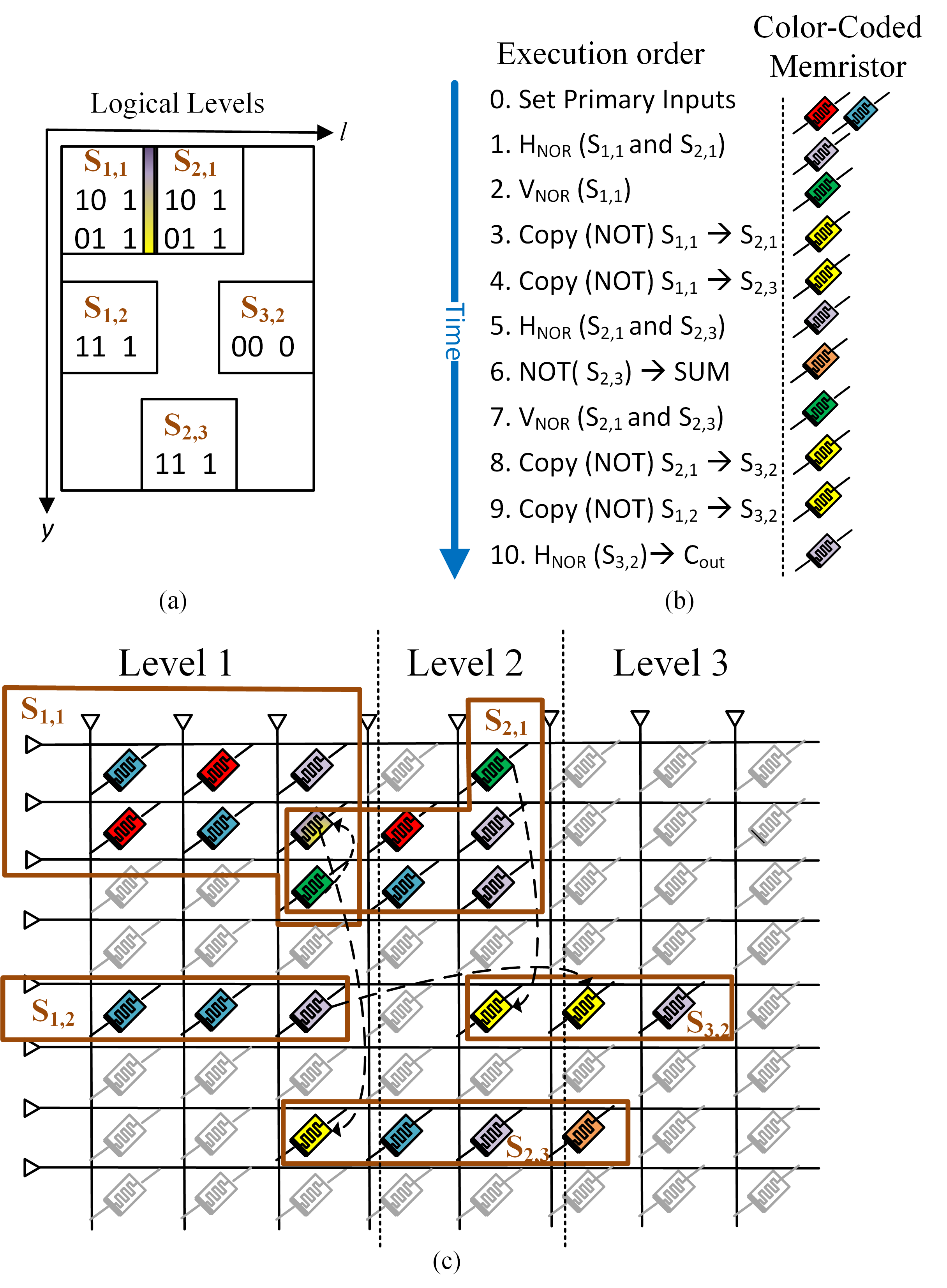}
\caption{\small Implemented 1-bit full adder using the HIPE-MAGIC: (a) tile-based mapping, (b) execution orders, and (c) MCA mapping. 10 cycles of PIM operations and 22 memristors are needed.}
\label{fig:proposed}
\end{figure}

\textcolor{black}{Using this placement scheme, supergates corresponding to $S_{2,3}$ and $S_{3,3}$ in Fig. \ref{fig:SAID} (a) are pushed up in their respective columns (i.e., $S_{2,1}$ and $S_{3,2}$ in Fig. \ref{fig:proposed} (a), respectively), which improves the area efficiency. Furthermore, the output memristor of $S_{1,1}$ is shared as one of inputs of $S_{2,1}$, as they are cascaded supergates in the same \textit{y} position, which saves one data alignment copy. }\textcolor{black}{Note that the assumption for the underlying MCA dimension in this work is $1024 \times 1024$, which is a typical value range for dimensions of an MCA based on \cite{bitlet}. This means that the total number of terms within supergates of one logical level must not exceed 1024, which was the case in our experiments.}

\textbf{Resource Sharing for Data Alignment Copies: } 
When the output of a supergate should be aligned to any row(s) of cascaded supergates in the next logic level, two NOT copies are used for each alignment with the aid of one auxiliary memristor. However, by sharing the auxiliary memristor for alignment copies that must be carried out for the output of the supergate and doing the second NOT for each alignment from that memristor (represented by \textit{Share\_Align\_Memristors()} function in Algorithm \ref{alg.mapping}), we reduce both latency and memristor count. \textcolor{black}{In Fig. \ref{fig:proposed}, we see an example for the advantage of resource sharing for data alignment copies, as we reuse one of input memristors of $S_{2,1}$ for transferring output of $S_{1,1}$ to $S_{2,3}$.}


\section{Experimental Results} \label{results}
We used ISCAS'85 and IWLS'93 benchmark suites for our performance evaluation. A Python script runs ABC by leveraging optimizing scripts during synthesis and varying $k$ values \{$k$ = 2,3,7,10\}. Next, technology mapping is done according to the mapping scheme presented in Section \ref{sec.mapping}. \replaced[id=ae]{\textcolor{black}{Lastly, the design spaces for each benchmark (associated with different values of $k$) is explored through a Pareto analysis and the best solution for each benchmark in terms of the area-latency trade-off is determined as the solution associated with the HIPE-MAGIC.}}{Lastly, the best solution in terms of the area-latency trade-off for each benchmark is determined through a Pareto analysis} \textcolor{black}{
Note that the real-time control of the MCA implementing the logic function (i.e., a sequencer which activates rows and columns of the MCA) is not considered.
} Also, it is worth mentioning that considering the electrical conditions of MCA such as current sneak path is out of the scope of this paper.

\begin{table*}[t]
   \caption{ \small Comparison between the mapping metrics of HIPE-MAGIC with those of \cite{Said} and \cite{SimAnn} on ISCAS'85 and IWLS'98 benchmarks.} 
   \label{tab:Results}
   \resizebox{\textwidth}{!}{%
   \centering 
   \begin{tabular}{c|c|cc|cccc|cccc} 
   \toprule[\heavyrulewidth]\toprule[\heavyrulewidth]
   \multirow{2}{*}{\textbf{Benchmark}}& \multirow{2}{*}{\textbf{Circuit}} & \multicolumn{2}{c|}{\textbf{HIPE-MAGIC}} & \multicolumn{4}{c|}{\textbf{\textcolor{black}{Tenace et al.} \cite{Said}}} & \multicolumn{4}{c}{\textbf{\textcolor{black}{Thangkhiew and Datta} \cite{SimAnn}}}  \\
   \cmidrule{3-12}
   &  &\textbf{Cycles} & \textbf{Mems} & \textbf{Cycles} & \textbf{Mems} & \textbf{Speedup} & \textbf{Area-Saving} & \textbf{Cycles} & \textbf{Mems} & \textbf{SpeedUp} & \textbf{Area-Saving} \\
   \midrule
   \multirow{11}{*}[-0.4ex]{\rotatebox[origin=c]{90}{\textbf{ISCAS'85}}}
   & \textbf{c432}   & 122 & 366   & 156  & 631  & 1.28 & 1.72      & 265  & 372  & 2.17 & 1.02 \\
   & \textbf{c499}   & 253 & 836   & 420  & 1399 & 1.66 & 1.67      & 935  & 1085 & 3.70 & 1.30 \\
   & \textbf{c880}   & 219 & 862   & 482  & 1113 & 2.20 & 1.29      & 750  & 886  & 3.42 & 1.03 \\
   & \textbf{c1355}  & 253 & 836   & 554  & 1182 & 2.19 & 1.41      & 938  & 1076 & 3.71 & 1.29 \\
   & \textbf{c1908}  & 313 & 809   & 627  & 1095 & 2.00 & 1.35      & 970  & 1061 & 3.10 & 1.31 \\
   & \textbf{c2670}  & 332 & 1462  & 643  & 1249 & 1.94 & 0.85      & 1401 & 1915 & 4.22 & 1.31 \\
   & \textbf{c3540}  & 758 & 2544  & 1566 & 3261 & 2.06 & 1.29      & 2418 & 2663 & 3.19 & 1.05 \\
   & \textbf{c5315}  & 1043 & 3556 & 1754 & 2937 & 1.68 & 0.83      & 3239 & 3795 & 3.11 & 1.07 \\
   & \textbf{c6288}  & 2429 & 5141 & 4069 & 6067 & 1.67 & 1.19      & 5007 & 5344 & 2.06 & 1.04 \\
   & \textbf{c7552}  & 1510 & 3507 & 2565 & 4003 & 1.69 & 1.14      & 3824 & 4415 & 2.53 & 1.26 \\ 
    \cmidrule{2-12} 
   & \textbf{Average} & \textbf{723.2} & \textbf{1991.9} & \textbf{1283.6} & \textbf{2293.7} & \textbf{1.84} & \textbf{1.27} & \textbf{1974.7} & \textbf{2261.2} & \textbf{3.12} & \textbf{1.17} \\
   \midrule
   \multirow{10}{*}[-0.4ex]{\rotatebox[origin=c]{90}{\textbf{IWLS'98}}} 
   & \textbf{9sym}     & 99  & 154  & 160 & 1026 & 1.61 & 1.78 & 552  & 594  & 5.57  & 1.03 \\
   & \textbf{apex5}    & 371 & 1910 & 777 & 2223 & 2.09 & 1.16 & 1966 & 2319 & 5.30  & 1.21 \\
   & \textbf{clip}     & 54  & 310  & 135 & 451  & 2.50 & 1.45 & 239  & 261  & 4.42  & 0.84 \\
   & \textbf{duke2}    & 263 & 1037 & 300 & 1632 & 1.14 & 1.57 & 1261 & 1342 & 4.80  & 1.29 \\
   & \textbf{inc}      & 21  & 296  & 55  & 280  & 2.62 & 0.95 & 264  & 282  & 12.57 & 0.95 \\
   & \textbf{misex3c}  & 183 & 1585 & 518 & 2551 & 2.83 & 1.61 & 2976 & 3094 & 16.26 & 1.95 \\
   & \textbf{rd73}     & 27  & 232  & 150 & 379  & 5.55 & 1.63 & 262  & 280  & 9.70  & 1.21 \\
   & \textbf{sao2}     & 54  & 344  & 79  & 559  & 1.46 & 1.62 & 309  & 331  & 5.72  & 0.96 \\
   & \textbf{vg2}      & 27  & 172  & 55  & 280  & 2.04 & 1.63 & 289  & 345  & 10.70 & 2.01 \\
    \cmidrule{2-12}
   & \textbf{Average} & \textbf{122.11} & \textbf{671.11} & \textbf{247.67} & \textbf{1042.33} & \textbf{2.43} & \textbf{2.03} & \textbf{902} & \textbf{983.11} & \textbf{8.34} & \textbf{1.59} \\
   \bottomrule[\heavyrulewidth] 
   \end{tabular}}
\end{table*}

\subsection{Operation Complexity and Area Comparison}
We evaluate the proposed approach by calculating the numbers of compute cycles and memristors required for each benchmark using the proposed synthesis and mapping flow. Area and latency improvements of the HIPE-MAGIC with respect to some of the state-of-the-art prior work are reported in TABLE \ref{tab:Results}. 
We compare HIPE-MAGIC with the mapping approach presented in \cite{Said}, and the work in \cite{SimAnn} which presents a mapping procedure for a netlist consisting of MAGIC NOR/NOT gates based on a simulated annealing algorithm \added[id=af]{to minimize the number of computational cycles}. For each MCA-based implementation, we compare the number of cycles required to generate all primary outputs (Cycles in TABLE \ref{tab:Results}), and the number of memristors required to implement the synthesized netlist (Mems in TABLE \ref{tab:Results}).\deleted[id=ae]{Since all memristors are of equal size, comparing the number of memristors for each circuit ensures a fair area comparison.}
\deleted[id=ae]{To evaluate the area-efficiency (AreaEff in TABLE \ref{tab:Results}), we use the following metric which reflects the density of active memristors with respect to the MCA size:}
\deleted[id=ae]{In Table \ref{tab:Results}, we only include those IWLS'98 benchmarks for which their area-efficiency figure employing the mapping procedure of \cite{Said}, has been reported in \cite{Simpler}).}

Results confirm that proposed optimization scripts leveraging the balancing operations reduce the latency by offering more opportunities for executing multiple supergates of a logic level concurrently, while simultaneously reducing the number of logic levels. \textcolor{black}{ However, the proposed scripts may add some redundancies in terms of logic and/or supergates to keep the logic level balanced, which may introduce area overheads. A particular mention is required for c2670 and c5315 where HIPE-MAGIC has a small area overhead compared to \cite{Said}. It is worth mentioning that a favorable trade-off between
speed and area can be obtained in HIPE-MAGIC by setting a different value of \textit{k}.} In addition, the dedicated proposed mapping flow compensates the potential area overhead due to redundancy introduced by supergates, and also reduces the latency mainly by sharing resources for data alignment copies. In particular, HIPE-MAGIC outperforms \cite{Said} in terms of the latency by an average of 2.11x improvement across both benchmark suites, while using 1.37x fewer memristors. Concerning the meta-heuristic mapping solution presented in \cite{SimAnn}, HIPE-MAGIC is 3.12x (8.34x) faster, requiring 1.17x (1.59x) fewer memristors for ISCAS'85 (IWLS'98) benchmarks. 

\deleted[id=af]{Notice that the mapping procedure of \cite{Said} does not achieve any improvements for \textbf{c6288}, because this circuit is an array multiplier that uses many full adders (see Fig. \ref{fig:SAID} (c)). HIPE-MAGIC can efficiently map this benchmark with metrics close to those of \cite{MagicBenefits} \textcolor{black}{why \cite{MagicBenefits}?} demonstrating the scalability and generality of HIPE-MAGIC approach.}
\deleted[id=ae]{
Finally, it is worth mentioning that while HIPE-MAGIC obtains better results in terms of area efficiency \textbf{shouldn't be area saving?} by densely placing supergates in each column, the improvement is still marginal as the number of LUTs for different columns could vary which leads to not fully taking advantage of the underlying 2D grid.
}


\subsection{\textcolor{black}{Analytical Comparison - PIM vs. CPU}} \label{analytical_compare}
\textcolor{black}{We analyze and quantify the strengths and weaknesses of our PIM realization over conventional CPU and other PIM solutions from [17] and [14].} \textcolor{black}{For our comparison, we employ a parameterized analytical modeling tool presented in \cite{bitlet} (and referred to as \textit{Bitlet}), which evaluates the affinity of workloads to PIM versus CPU computing according to some PIM- and CPU-related parameters.
PIM computations in the Bitlet model are limited mainly by operation complexity and data alignment costs, i.e., the number of PIM cycles in Table \ref{tab:Results}. On the other hand, for CPU computing, Bitlet ignores the cost associated with computations and data movements done within the CPU itself (which is indeed in the favor of CPU computing for our comparison), and the CPU throughput is thus mainly limited by its  external memory bandwidth utilization, i.e., the cost associated with the number of data bits transferred between the CPU and memory \cite{bitlet}.} 
\textcolor{black}{Bitlet uses technology parameters for a 28nm node from \cite{CPU_DRAM} for the memory model in CPU and bases its model of memristor on \cite{memmodel}.}

\begin{figure}[t]
     \centering
     \begin{subfigure}[b]{0.4\textwidth}
         \centering
         \includegraphics[width=\textwidth]{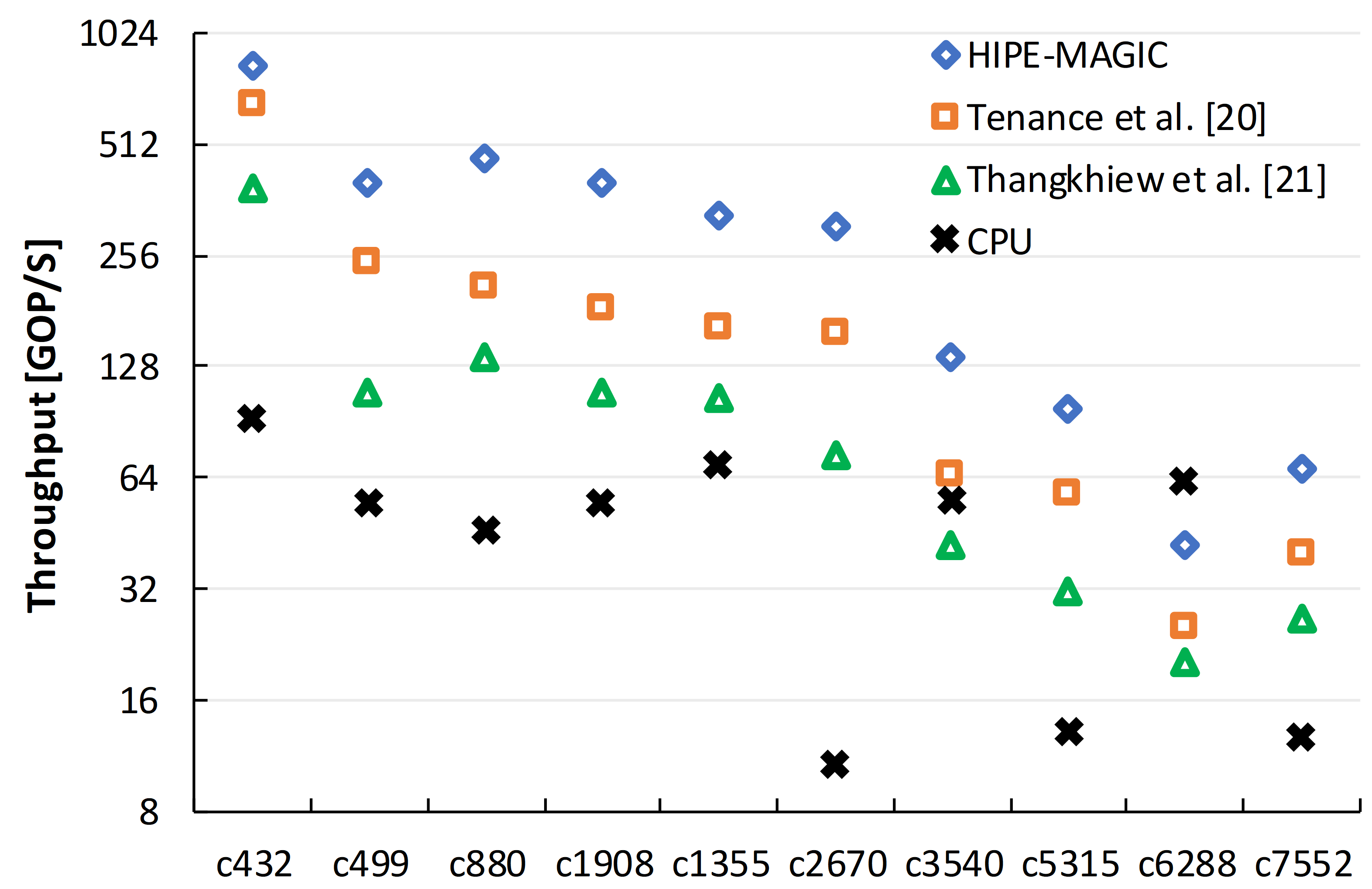}
         \caption{}
          \vspace{5mm}
         \label{fig:ThEval}
     \end{subfigure}
     \hfill
     \begin{subfigure}[b]{0.4\textwidth}
         \centering
         \includegraphics[width=\textwidth]{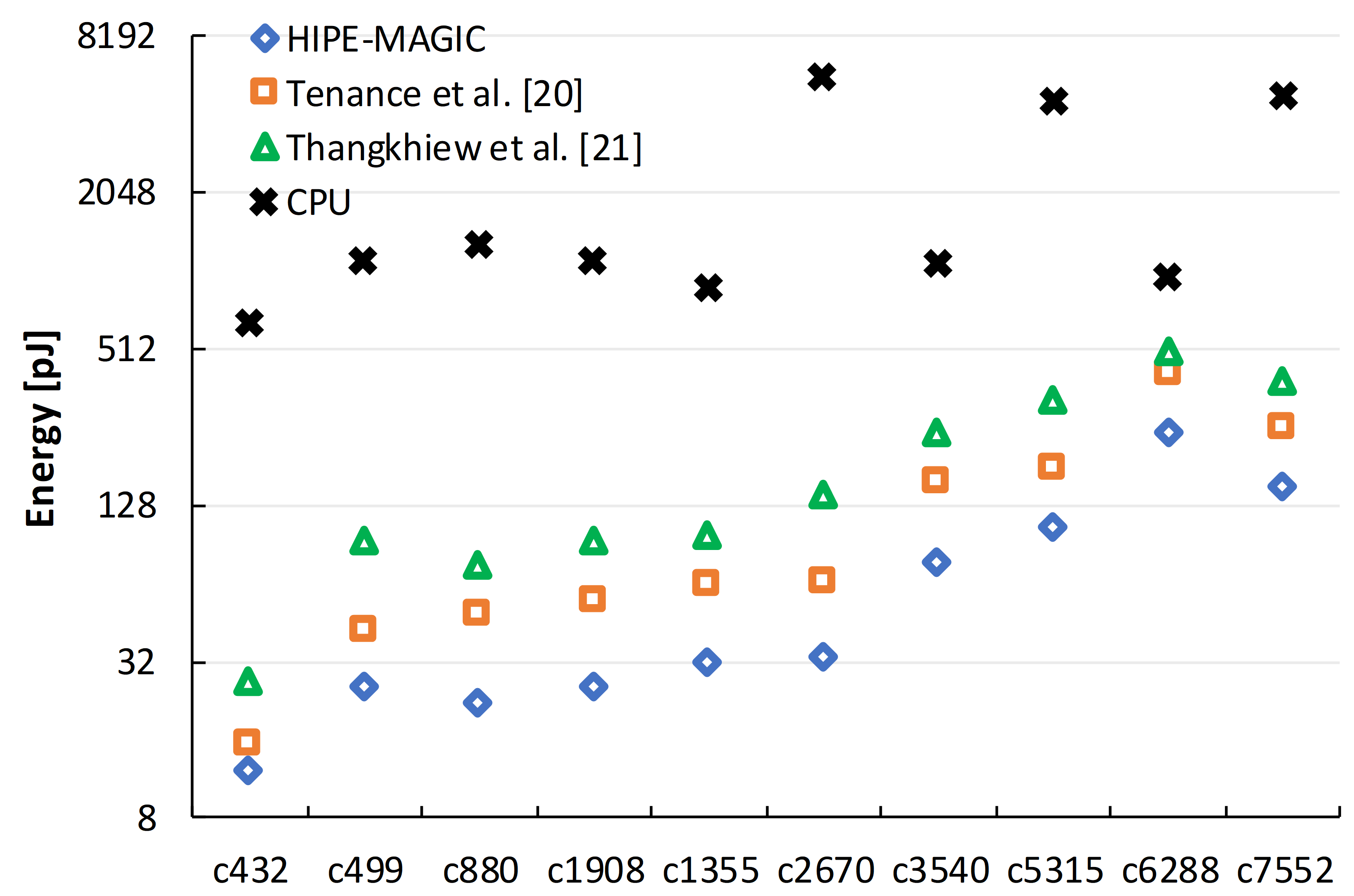}
         \caption{}
         \label{fig:EnEval}
     \end{subfigure}
        \vspace{5mm}
        \caption{\textcolor{black}{\small Throughput and energy consumption comparisons between CPU and PIM approaches.}}
        \vspace{3mm}
        \label{fig:eval}
\end{figure}

\textcolor{black}{Using the Bitlet model,
Fig. \ref{fig:eval} (a) shows a comparison in terms of achieved throughput between PIM solutions (including HIPE-MAGIC) and CPU computing.}
\textcolor{black}{Notice that the PIM does not achieve any improvements in terms of throughput for c6288, because this circuit is an array multiplier with a relatively large number of PIM cycles. Furthermore, the
comparison between PIM and CPU with respect to c2670, which has a relatively large number of input/output lines (which affects the CPU throughput adversely)}, demonstrates that PIM can greatly alleviate the overheads associated with data transfers from/to the main memory. 

\textcolor{black}{Comparing the energy obtained using the Bitlet model, CPU energy consumption is higher than that of PIM solutions over all the studied benchmarks (see Fig. \ref{fig:eval} (b)). For the CPU, energy consumption is associated with the number of transferred bits and energy per bit transfer, while operation complexity and PIM energy per each operation determine the energy consumption for PIM solutions. As the number of PIM cycles increase, the relative energy efficiency of PIM decreases (e.g., c6288).} 

These results demonstrate once again that PIM-based solutions are beneficial if much of the latency and energy consumption associated with a target application is due to data transfers between memory and the computing fabric. There are many such interesting applications including for example big data analysis, neural network based inference, and so on.

\section{Conclusion}\label{conc}
In this paper, we presented a technology-aware synthesis approach along with a heuristic mapping framework. 
\textcolor{black}{Results show that the HIPE-MAGIC achieves promising results compared to  state-of-the-art work and the conventional CPU computing. As data alignment copy operations contribute to significant number of cycles for implementing a logic function, restricting the number of fan-outs of LUT nodes during logic synthesis can be a promising future work.}

\balance
\bibliographystyle{ACM-Reference-Format}
\bibliography{sample-base}


\end{document}